\documentstyle [11pt,moriond,epsfig]{article}

\def\Jnl #1#2#3#4 {#1 {\bf #2}, (#3) #4}
\def\NPB {{\em Nucl. Phys.} {\bf B}}
\def\PLB {{\em Phys. Lett.}  {\bf B}}

\def\PRD {{\em Phys. Rev.} {\bf D}}

\def\EPJC {{\em Eur. Phys. J.} {\bf C}}
\def\lsim {\;\raise0.3ex\hbox{$<$\kern-0.75em\raise-1.1ex\hbox{$\sim$}}\;}
\def\gsim {\;\raise0.3ex\hbox{$>$\kern-0.75em\raise-1.1ex\hbox{$\sim$}}\;}
\def\ZZ {\hbox{\it Z\hskip -4.pt Z}}
\def\wt{\widetilde}
\def\be {\begin{equation}}
\def\ee {\end{equation}}
\def\ba {\begin{array}}
\def\ea {\end{array}}
\def\bea {\begin{eqnarray}}
\def\eea {\end{eqnarray}} 
\def\ben{\begin{enumerate}} 
\def\een{\end{enumerate}}
\def\nn {\nonumber}
\def\ni {\noindent}        

\begin {document}

\ni RAL-TR-2000-027\\

\vspace* {4cm}
\title {HIGGS MASSES AND COUPLINGS IN THE NMSSM}

\author {U. ELLWANGER$^a$, \underline {C. HUGONIE}$^b$}

\address { $^a$ Laboratoire de Physique Th\'eorique,\\
Universit\'e de Paris XI, F-91405 Orsay Cedex, France\\
$^b$ Rutherford Appleton Laboratory,\\
Chilton, Didcot, OX11 0QX, England}

\maketitle
\abstracts {We give the upper bounds on the masses of the lightest and second
lightest CP even Higgs bosons in the NMSSM, the MSSM extended by a gauge
singlet. The dominant two loop corrections are included. Since the coupling $R$
of the lightest Higgs scalar to gauge bosons can be small, we study in detail
the relation between masses and couplings of both lightest scalars. We present
upper bounds on the mass of a 'strongly' coupled Higgs ($R > 1/2$) as a function
of lower experimental limits on the mass of a 'weakly' coupled Higgs ($R <
1/2$). With the help of these results, the whole parameter space of the model
can be covered.}

\section {Introduction}

The NMSSM \cite {NMSSM} (Next-to-Minimal Supersymmetric Standard Model, also
called (M+1)SSM) is defined by the addition of a gauge singlet superfield $S$
to the MSSM and a $\ZZ_3$ symmetry of the renormalizable part of the
superpotential $W$. It allows to omit the so-called $\mu$ term $\mu H_1 H_2$ in
the superpotential of the MSSM, and to replace it by a Yukawa coupling (plus a
singlet self coupling), hence solving the $\mu$ problem of the MSSM. Apart from
the standard quark and lepton Yukawa couplings, $W$ is given by
\be
W = \lambda H_1 H_2 S + {1\over3} \kappa S^3 + ... \label {supot}
\ee
and the corresponding trilinear couplings $A_\lambda$ and $A_\kappa$ are
added to the soft susy breaking terms. Once the electroweak symmetry is broken,
the scalar component of $S$ acquires a vev $s = \langle S \rangle$, thus
generating an effective $\mu$ term with $\mu = \lambda s$. The superpotential
(\ref {supot}) is scale invariant, and the electroweak scale appears only
through the susy breaking terms. The possible domain wall problem due to the
discrete $\ZZ_3$ symmetry is assumed to be solved by adding non-renormalizable
interactions which break the $\ZZ_3$ symmetry without spoiling the quantum
stability \cite {walls}.

The new physical states of the NMSSM are one additional neutral Higgs scalar
and one Higgs pseudoscalar, respectively, and one additional neutralino.
Therefore, the Higgs sector of the model consists in 3 scalar states denoted as
$S_i$ with masses $m_i$, $i=1..3$, in increasing order, and 2 pseudoscalar
states denoted as $P_i$ with masses $m'_i$, $i=1,2$, in increasing order.

In view of ongoing Higgs searches at LEP2 \cite {LEP2} and, in the near future,
at Tevatron Run II \cite {Teva}, it is important to check the model dependence of
upper bounds on Higgs masses. Within the MSSM, the mass of the lightest CP even
Higgs boson is bounded, at tree level, by
\be
m_h^2 \leq M_Z^2 \cos^2 2\beta .
\ee
It has been realized already some time ago that loop corrections weaken
this upper bound. This corrections depend on the soft susy breaking terms of
$O(M_{susy})$. At the one loop level, assuming $M_{susy} \leq$ 1 TeV, the upper
limit on $m_h$ is $\sim$ 140 GeV. Also two loop corrections have been
considered in the MSSM \cite {2loop}; these have the tendency to lower the
upper bound on $m_h$ by at most $\sim$ 10 GeV.

The aim of these proceedings is to study the upper limits on Higgs masses
including two loop corrections in the NMSSM. All the results displayed here
were originally presented in ref.\cite {paper}.

\section {Upper bound on the lightest Higgs mass}

In the NMSSM, the upper bound on the mass $m_1$ of the lightest CP even Higgs
differs from the one of the MSSM already at tree level:
\be
m_1^2 \leq M_Z^2 \left ( \cos^2 \! 2\beta + \frac {2\lambda^2}{g_1^2+g_2^2} \,
\sin^2 \! 2\beta \right ) \label {tree}
\ee
where $g_1$ and $g_2$ denote the $U(1)_Y$ and the $SU(2)_L$ gauge
couplings. Note that, for $\lambda < .53$, $m_1$ is still bounded by $M_Z$ at
tree level. Large values of $\lambda$ are in any case prohibited, if one
requires the absence of a Landau singularity for $\lambda$ below the GUT scale.
The upper bound on $\lambda$ at the weak scale depends on the value of $\kappa$
and on the top quark Yukawa coupling $h_t$, i.e. on $\tan\!\beta$ (cf. fig.
1).

\begin {figure}[ht]
\begin {center}
\epsfig {file=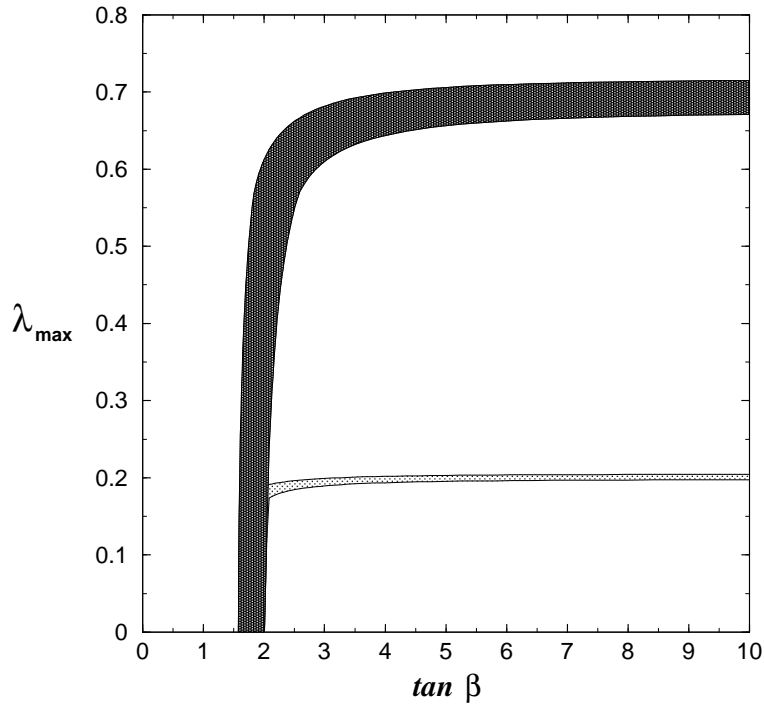,width=10cm}
\end {center}
\caption {Upper bound on $\lambda$ as a function of $\tan\!\beta$ for
$m_t^{pole} = 173.8 \pm 5.2$ GeV, $\kappa = 0$ (dark) and $\kappa = 0$
(light).}
\end {figure}

In order to obtain the correct upper limits on the Higgs boson masses in the
presence of soft susy breaking terms, radiative corrections to several terms in
the effective action have to be considered. Let us first introduce a scale $Q
\sim M_{susy}$, where $M_{susy}$ is of the order of the susy breaking terms.
Let us assume that quantum corrections involving momenta $p^2 \gsim Q^2$ have
been evaluated; the resulting effective action $\Gamma_{ef\!f}(Q)$ is then
still of the standard supersymmetric form plus soft susy breaking terms.
One is left with the computation of quantum corrections to $\Gamma_{ef\!f}$
involving momenta $p^2 \lsim Q^2$. Subsequently the quantum corrections to the
following terms in $\Gamma_{ef\!f}$ will play a role:

\ben
\item[a)] Corrections to the Higgs effective potential. The effective potential
$V_{ef\!f}$ can be developped in power of $\hbar$ or loops as
\be
V_{ef\!f} = V^{(0)} + V^{(1)} + V^{(2)} + \ldots .
\ee
The tree level potential $V^{(0)}$ is determined by the usual F and D terms
and the standard soft susy breaking terms. The one loop corrections to the
effective potential are given by
\be
V^{(1)} = \frac {1}{64\pi^2} \, \mbox {STr} M^4 \left [ \ln \left (
\frac {M^2}{Q^2} \right ) - \frac {3}{2} \right ] , \label {V1}
\ee
where we only take top and stop loops into account. Next, we consider the
dominant two loop corrections. These will be numerically important only for
large susy breaking terms compared to the Higgs vevs $h_i$, hence we can expand
in powers of $h_i$. Since the terms quadratic in $h_i$ can be absorbed into the
tree level soft terms, we just consider the quartic terms, and here only those
which are proportional to large couplings: terms $\sim \alpha_s h_t^4$ and
$\sim h_t^6$. Finally, we are only interested in leading logs. The
corresponding expression for $V^{(2)}$ is
\be
V^{(2)}_{LL} = 3 \left ( \frac {h_t^2}{16\pi^2} \right )^2 h_1^4 \left (
32\pi\alpha_s - \frac {3}{2} \, h_t^2 \right ) t^2 \label {V2}
\ee
where $t \equiv \ln \left (\frac{M^2}{Q^2} \right )$ and $h_1 = \langle H_1
\rangle$, $H_1$ being, in our conventions, the Higgs which couples to the top quark.

\item[b)] Corrections to the kinetic terms of the Higgs bosons. They lead to a
wave function renormalization factor $Z_{H_1}$ in front of the $D_\mu H_1 D^\mu
H_1$ term with, to order $h_t^2$,
\be
Z_{H_1} = 1 + 3 \frac {h_t^2}{16\pi^2} \, t . \label {zh}
\ee
Due to gauge invariance the same quantum corrections contribute to the kinetic
energy and to the Higgs-$Z$ boson couplings, which affect the relation between
the Higgs vevs and $M_Z$.

\item[c)] Corrections to the Higgs-top quark Yukawa coupling. After an
appropriate rescaling of the $H_1$ and top quark fields in order to render
their kinetic terms properly normalized, these quantum corrections lead to an
effective coupling $h_t(m_t)$ with, to orders $h_t^2$, $\alpha_s$,
\be
h_t(m_t) = h_t(Q) \left ( 1 + \frac {1}{32\pi^2} \left ( 32\pi\alpha_s -
\frac {9}{2} \, h_t^2 \right ) t \right ) . \label {htmt}
\ee
The (running) top quark mass is then given by
\be
m_t(m_t) = h_t(m_t) Z_{H_1}^{1/2} h_1 \label {mtrun}
\ee
and the relation between the pole and running mass, to order $\alpha_s$,
reads
\be
m_t^{pole} = m_t(m_t) \left ( 1 + \frac {4\alpha_s}{3\pi} \right ) .
\label{mtpole}
\ee

\een

Taking into account these corrections and assuming $h_i \ll M_{susy}$, one
obtains the following upper limit on the lightest CP even Higgs mass:
\bea
m_1^2 & \leq &  M_Z^2 \left ( \cos^2 \! 2\beta + \frac {2\lambda^2}{g_1^2+g_2^2}
\, \sin^2 \! 2\beta \right ) \left ( 1 - \frac {3h_t^2}{8\pi^2} \, t \right )
\nn \\
& & + \frac {3h_t^2(m_t)}{4\pi^2} \, m_t^2(m_t) \sin^2 \! \beta \left ( \frac
{1}{2} \, \wt{X}_t + t + \frac {1}{16\pi^2} \left ( \frac {3}{2} \, h_t^2 -
32\pi\alpha_s \right ) ( \wt{X}_t + t ) t \right ) \label {below1}
\eea
\bea
\mbox {where} & & \wt{X}_t \equiv 2 \, \frac {\wt{A}_t^2}{M_{susy}^2} \left ( 1
- \frac {\wt{A}_t^2}{12M_{susy}^2} \right ) \label {Xt}\\
\mbox {and} & & \wt{A}_t \equiv A_t - \lambda s \cot\beta ,
\eea
$A_t$ being the top trilinear soft term and $s$ the vev of the singlet.

\begin {figure}[hb]
\begin {center}
\epsfig {file=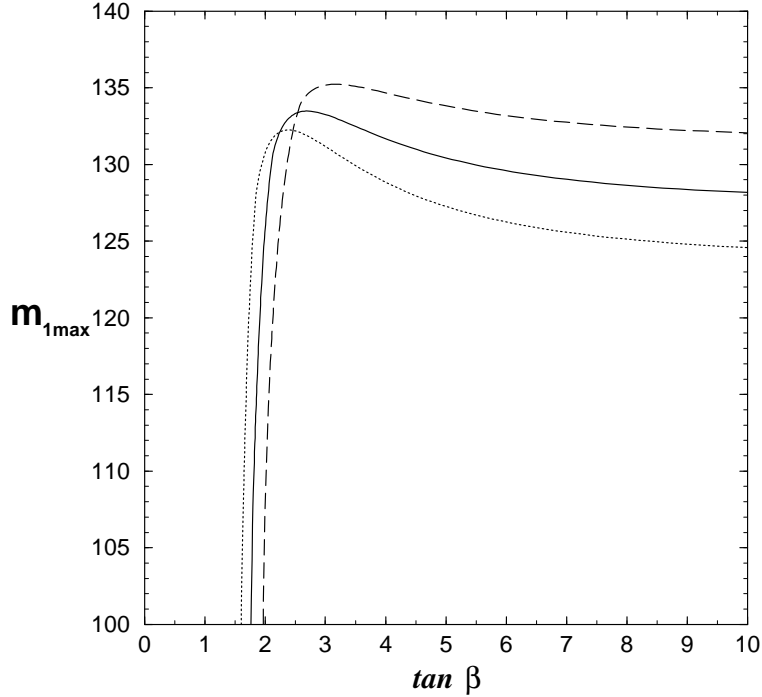,width=10cm}
\end {center}
\caption {Upper bound on $m_1$ [GeV] versus $\tan\beta$ for $m_t^{pole} =
173.8 \pm 5.2$ GeV (straight, dashed, dotted line respectively), and $M_{susy}
\leq 1$ TeV.}
\end {figure}

\ni The only difference between the MSSM bound \cite{2loop} and (\ref {below1})
is the 'tree level' term $\sim \lambda^2 \sin^2 \! 2\beta$. This term is
important for moderate values of $\tan\!\beta$. Hence, the maximum of the
lightest Higgs mass in the NMSSM is not obtained for large $\tan\!\beta$ as in
the MSSM, but rather for moderate $\tan\!\beta$ (cf. fig. 2). On the other
hand, the radiative corrections are identical in the NMSSM and in the MSSM. In
particular, the linear dependence in $\wt{X}_t$ is the same in both models.
Hence, from eq.~(\ref {Xt}), the upper bound on $m_1^2$ is maximized for
$\wt{X}_t=6$ (corresponding to $\wt{A}_t=\sqrt{6} M_{susy}$, the 'maximal
mixing' case), and minimized for $\wt{X}_t=0$ (corresponding to $\wt{A}_t=0$,
the 'no mixing' case).

\section {Mass bounds versus reduced couplings}

However, the upper limit on $m_1$ is not necessarily physically relevant, since
the coupling of the lightest CP even Higgs to the $Z$ boson can be very
small. Actually, this phenomenon can also appear in the MSSM, if
$\sin^2(\beta-\alpha)$ is small. However, the CP odd Higgs boson $A$ is then
necessarily light ($m_A \sim m_h < M_Z$ at tree level), and the process $Z
\rightarrow h A$ can be used to cover this region of the parameter space in the
MSSM. In the NMSSM, a small gauge boson coupling of the lightest Higgs $S_1$
is usually related to a large singlet component, in which case no
(strongly coupled) light CP odd Higgs boson is available. Hence, Higgs searches
in the NMSSM have possibly to rely on the search for the second lightest
Higgs scalar $S_2$.

Let us now define the reduced coupling $R_i$ as the square of the coupling
$ZZS_i$ divided by the corresponding standard model Higgs coupling:
\be
R_i = (S_{i1}\sin\!\beta + S_{i2}\cos\!\beta)^2
\ee
where $S_{i1}, S_{i2}$ are the $H_1, H_2$ components of the CP even Higgs
boson $S_i$, respectively. Evidently, we have $0 \leq R_i \leq 1$ and unitarity
implies
\be
\sum_{i=1}^3 R_i = 1 . \label {Rsum}
\ee

We are interested in upper limits on the two lightest CP even Higgs bosons
$S_{1,2}$. These are obtained in the limit where the third Higgs, $S_3$, is
heavy and decouples, i.e. $R_3 \sim 0$ (This is the equivalent of the so called
decoupling limit in the MSSM: the upper bound on the lightest Higgs $h$ is
saturated when the second Higgs $H$ is heavy and decouples). Hence, we have
$R_1 + R_2 \simeq 1$.

In the regime $R_1 \geq 1/2$ experiments will evidently first discover the
lightest Higgs (with $m_1 \leq 133.5$ GeV for $m_t^{pole} = 173.8$ GeV and
$M_{susy} = 1$ TeV). The 'worst case scenario' in this regime corresponds to
$m_1 \simeq 133.5$ GeV, $R_1 \simeq 1/2$; the presence of a Higgs boson with
these properties has to be excluded in order to test this part of the parameter
space of the NMSSM.

In the regime where $R_1 < 1/2$ (and hence $1/2 < R_2 \leq 1$) the lightest
Higgs may escape detection because of its small coupling, and it may be easier
to detect the second lightest Higgs. In fig. 3 we show the upper limit on $m_2$
as a function of $R_2$ as a thin straight line. For $R_2 \rightarrow 1$
(corresponding to $R_1 \rightarrow 0$) the upper limit on $m_2$ is actually
given by the previous upper limit on $m_1$, even if the corresponding Higgs
boson is the second lightest one. For $R_2 \rightarrow .5$, on the other hand,
$m_2$ can be as large as 190 GeV. However, one finds that the upper limit on
$m_2$ is saturated only when the mass $m_1$ of the lightest Higgs boson tends
to 0. Clearly, one has to take into account the constraints from Higgs boson
searches which apply to reduced couplings $R < 1/2$, i.e. lower limits on $m_1$
as a function of $R_1 \simeq 1 - R_2$, in order to obtain realistic upper
limits on $m_2$ vs. $R_2$. The dotted curves in fig. 3 show the upper limit on
$m_2$ as a function of $R_2$ for different fixed values of $m_1$ (as indicated
on each curve). They can be used to obtain upper limits on the mass $m_2$, in
the regime $R_1 < 1/2$, for arbitrary experimental lower limits on the mass
$m_1$: For each value of the coupling $R_1$, which would correspond to a
vertical line in fig. 3, one has to find the point where this vertical line
crosses the dotted curve associated to the corresponding experimental lower
limit on $m_1$. Joining these points by a curve leads to the upper limit on
$m_2$ as a function of $R_2$. We have indicated the present LEPII limit \cite
{LEP2}, which give, in the 'worst case' an upper limit on $m_2$ of $\simeq$ 160
GeV for $R_2 \simeq .5$.

\begin {figure}[ht]
\begin {center}
\epsfig {file=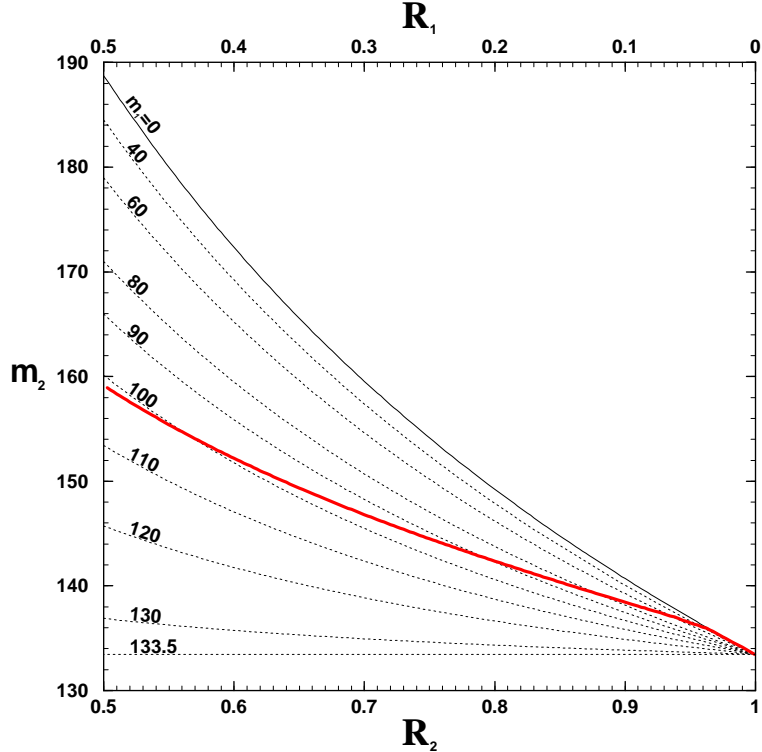,width=10cm}
\end {center}
\caption {Upper limits on the mass $m_2$ (in GeV) against $R_2$, for different
values of $m_1$ (as indicated on each line in GeV), assuming $m_t^{pole} =
173.8$ GeV and $M_{susy} \leq 1$ TeV. $R_1=1-R_2$ is shown on the the top axis.
The thick straight line corresponds to LEPII lower limits on $m_1$ vs. $R_1$.}
\end {figure}

Lower experimental limits on a Higgs boson with $R > 1/2$  restrict the allowed
regime for $m_2$ (for $R_2 > 1/2$) in fig. 3 from below. The present lower
limits on $m_2$ from LEP are not visible in fig. 3, since we have only shown
the range $m_2 > 130$ GeV. Possibly Higgs searches at Tevatron Run II will push
the lower limits on $m_2$ upwards into this range. This would be necessary if
one aims at an exclusion of the 'delicate' regime of the NMSSM: Then, lower
limits on the mass $m_2$ -- for any value of $R_2$ between $1/2$ and 1 -- of at
least 133.5 GeV are required; the precise experimental lower limits on $m_2$ as
a function of $R_2$, which would be needed to this end, will depend on the
achieved lower limits on $m_1$ as a function of $R_1$ in the regime $R_1 <
1/2$.

In principle, from eq. (\ref {Rsum}), one could have $R_2 > R_1$ with $R_2$ as
small as $1/3$. However, in the regime $1/3 < R_2 < 1/2$, the upper bound on
$m_2$ as a function of $R_2$ for different fixed values of $m_1$ can only be
saturated if $R_1 = R_2$. It is then sufficient to look for a Higgs boson with
a coupling $1/3 < R < 1/2$ and a mass $m \lsim 133.5$ GeV to cover this region
of the parameter space of the NMSSM.

\section {Conclusions}

We have emphasized the need to search for Higgs bosons with reduced couplings,
which are possible within the NMSSM. Our main results are presented in fig. 3,
which allows to obtain the constraints on the Higgs sector of the model both
from searches for Higgs bosons with weak coupling ($R < 1/2$), and strong
coupling ($R > 1/2$). The necessary (but not sufficient) condition for testing
the complete parameter space of the (M+1)SSM is to rule out a CP even Higgs
boson with a coupling $1/3 < R < 1$ and a mass below 135~GeV. The sufficient
condition (i.e. the precise upper bound on $m_2$ vs $R_2$) depends on the
achieved lower bound on the mass of a 'weakly' coupled Higgs (with $0 < R <
1/2$) and can be obtained from fig. 3. At the Tevatron this would probably
require an integrated luminosity of up to 30 fb$^{-1}$ \cite{Teva}. If this
cannot be achieved, and no Higgs is discovered, we will have to wait for the
results of the LHC in order to see whether supersymmetry beyond the MSSM is
realized in nature.

\section* {Acknowledgments}

C.H. would like to thank the organizers of the XXXVth Rencontres de Moriond for
the very stimulating atmosphere of this conference.


\section* {References}

\begin {thebibliography}{99}

\bibitem {NMSSM} H.P. Nilles, M. Srednicki, D. Wyler, \Jnl {\PLB}{120}{1983}{346};\\
J.M. Frere, D.R.T. Jones, S. Raby, \Jnl {\NPB}{222}{1983}{11};\\
J.P. Derendinger, C.A. Savoy, \Jnl {\NPB}{237}{1984}{307}.

\bibitem {walls} S.A. Abel, \Jnl {\NPB}{480}{1996}{55};\\
C. Panagiotakopoulos, K. Tamvakis, \Jnl {\PLB}{446}{1999}{224}.

\bibitem {LEP2} The LEP working group for Higgs boson searches, contribution to
these proceedings.

\bibitem {Teva} Physics at Tevatron Run II Workshop, Higgs working group final
report (to appear).

\bibitem {2loop} J. Espinosa, M. Quiros, \Jnl {\PLB}{266}{1991}{389};\\
R. Hempfling, A. Hoang, \Jnl {\PLB}{331}{1994}{99};\\
M. Carena, J. Espinosa, M. Quiros, C. Wagner, \Jnl {\PLB}{355}{1995}{209};\\
M. Carena, M. Quiros, C. Wagner, \Jnl {NPB}{461}{1996}{407};\\
S. Heinemeyer, W. Hollik, G. Weiglein, \Jnl {\PLB}{440}{1998}{296},
\Jnl {\PRD}{58}{1998}{091701}, \Jnl {\EPJC}{9}{1999}{343}, 
\Jnl {PLB}{455}{1999}{179};\\
R.-J. Zhang, \Jnl {\PLB}{447}{1999}{89}.

\bibitem {paper} U. Ellwanger, C. Hugonie hep-ph/9909260 (to appear in \EPJC).


\end {thebibliography}
\end {document}